\documentclass[twocolumn]{article}

\usepackage{amsmath}
\usepackage{booktabs}

\usepackage{graphicx,natbib}
\usepackage{authblk}
\usepackage{sectsty}
\sectionfont{\normalfont\large\bf}
\subsectionfont{\normalfont\it}

\def\bb{{\bf x}}
\def\bO{{\bf \Omega}}
\def\pcoh{\Phi}
\def\pha{\phi}
\def\rph{\alpha}
\def\vis{V}
\def\gobs{g}
\def\SNR{\hbox{\large$\epsilon$}}
\def\SNR{h}
\def\sigf{f}

\begin{document} 

\title{The theory of intensity interferometry revisited}

\author[1]{Prasenjit Saha}

\affil[1]{\small{\em Physik-Institut, University of Zurich,
  Winterthurerstrasse~190, 8057 Zurich, Switzerland}}

\setbox0=\vbox{\hsize=6.5truein
  \hbox to \hsize{\hss\bf Abstract\hss}
  \smallskip \noindent With the current revival of
  interest in astronomical intensity interferometry, it is interesting
  to revisit the associated theory, which was developed in the 1950s
  and 1960s.  This paper argues that intensity interferometry can be
  understood as an extension of Fraunhofer diffraction to incoherent
  light.  Interference patterns are still produced, but they are
  speckle-like and transient, changing on a time scale of
  $1/\Delta\nu$ (where $\Delta\nu$ is the frequency bandwidth) known
  as the coherence time.  Bright fringes average less than one photon
  per coherence time, hence fringes change before they can be
  observed.  But very occasionally, two or even more photons may be
  detected from an interference pattern within a coherence time.
  These rare coincident photons provide information about the
  underlying transient interference pattern, and hence about the
  source brightness distribution.  Thinking in terms of transient
  sub-photon interference patterns makes it easy to see why intensity
  interferometry will have large optical-path tolerance, and be immune
  to atmospheric seeing.  The unusual signal-to-noise properties also
  become evident.  We illustrate the unobservable but conceptually
  useful transient interference patterns, and their observable
  correlation signal, with three simulated examples: (i)~an elongated
  source like Achernar, (ii)~a three-star system like Algol, and
  (iii)~a crescent source that roughly mimics an exoplanet transit or
  perhaps the M87 black hole environment.  Of these, (i)~and (ii) are
  good targets for currently-planned setups, while (iii)~is
  interesting to think about for the longer term.}

\date{\box0}


\maketitle

\section{Introduction}\label{sec:intro}

The highest-resolution images known so far are all at radio
wavelengths.  The shadow of the M87 black hole
\citep{2019ApJ...875L...1E} is the best-known example, but the imaging
of BL~Lacertae by \cite{2016ApJ...817...96G} is nearly as remarkable.
The resolution in both cases is about $20\rm\,\mu as$ or
$10^{-10}$~radians, requiring interferometric baselines of
$\sim10^{10}\,\lambda$.  To match such resolution at optical
wavelengths would need baselines of several km.  Optical and infrared
interferometers \citep[e.g.,][]{2019A&A...628A.101H} have so far
achieved only a few hundred metres, because maintaining coherence over
longer baselines is very difficult.

A very different prospect for $10^{-10}$~radian resolution in visible
light is intensity interferometry. This is a method for imaging by
detecting coincident photons, without mirrors or lenses.  In practice,
mirrors are used for collecting light, but they are ordinary mirrors
and not of optical quality.  The usual principle in imaging
telescopes, that optical paths must be precise to sub-wavelength
tolerances, does not apply.  Instead, the optical-path tolerance is
set by the time-resolution of the photon counters, and is orders of
magnitude larger than the wavelength.  This makes intensity
interferometers relatively easy to build, and also immune to
atmospheric seeing.  The technical challenges have to do with
detecting and correlating photons.

The historic Narrabri Stellar Intensity Interferometer \citep[or NSII,
  see][]{1974iiaa.book.....H} observed with baselines to $180\rm\,m$,
corresponding to $0.5\rm\,mas$, with blue light.  With the
photon-detectors available in the 1960s, however, only 32 targets (all
hot stars) had enough SNR to resolve their angular sizes
\citep{1974MNRAS.167..121H}.  Intensity interferometry was then
abandoned.  In the last few years, there have been ambitious proposals
\citep{2013APh....43..331D,2014JKAS...47..235T} followed by new
measurements
\citep[e.g.,][]{2016SPIE.9907E..0NZ,2018SPIE10701E..0XW,2020MNRAS.491.1540A,2020MNRAS.494..218R,Abeysekara2020}
and further proposals \citep{2019BAAS...51g.227K}.

Intensity interferometry was not forgotten in the intervening decades.
The experiments of \cite{1956Natur.177...27B} are considered
foundational in quantum optics, and the HBT phenomenon has since been
observed with massive particles ---atoms and nucleii--- as well
\citep[for pointers to these areas, see][]{2008PhT....61h...8K}.  This
unusual history has led to two distinct sources for the theory: the
1950s development in astronomy
\citep{1954PMag...45..663B,1957RSPSA.242..300B}, and the 1960s
development in quantum optics \citep[e.g.,][]{1965RvMP...37..231M}
which modern optics books follow
\citep[e.g.,][]{Goodman2015,OuZhe-YuJeff2017Qofe}.  But neither of
these approaches is especially intuitive.  There is no nice example
one could point to, that illustrates the physics of HBT in a simple
way, and yet extends to more general cases.  In this paper we will
start by identifying a toy example, one that was not available in the
1960s, and use the intuition gained from it to help understand
intensity interferometry more simply.

Let us consider a double-slit interference experiment with
single-photon counting.  \cite{2013AmJPh..81..951R} present one nice
setup for this experiment, designed for physics teaching and
incorporating a few additional options.  Photons are detected at
random locations, but preferentially at bright fringes, building up
the fringe pattern with time.  This experiment illustrates that light
propagates as a wave but is detected as particles, a notion that is
standard now, but as is evident from \cite{1964PPS....84..435M}, was
still being debated when the theory of HBT was developed.  We now
imagine modifying the experiment, to something not far from recent
laboratory experiments on intensity interferometry
\citep[][]{2015A&A...580A..99D,2018JMOp...65.1336M,2020OExpr..28.5248Z}.
\begin{enumerate}
\item Introduce a slow random phase modulation at each slit.  This
  will make the interference pattern slide back and forth on the
  screen.
\item Next, split the screen and move part of it further from the
  slits, the pattern on that part will slide with a delay according to
  the light travel time.  It is as if the intensity pattern propagates
  away from the slits, and illuminates whatever screen it finds in its
  path.
\item Now reduce the intensity till the photons arrive less frequently
  than the phase modulation.  The fringes get washed out, because
  before even two photons are detected on a bright fringe, the fringe
  pattern will have moved.  Still, because of shot noise, sometimes
  two photons will arrive almost simultaneously.  Such photons will
  tend to be spaced by a multiple of the fringe width.  This feature
  will apply even across the split screen, provided we include the
  light travel time in our definition of simultaneous.  By confining
  our attention to these nearly-simultaneous photon pairs, we can
  infer the fringe width.
\item Finally, replace the slits with two small and narrow-band
  incoherent lights.  The beating of slightly different wavelengths in
  each light automatically produces random phase (and amplitude)
  modulation.
\end{enumerate}
We will consider a more general version of the above scenario in
Section~\ref{sec:transient} below.  In place of the double slit, we
start with Fraunhofer diffraction from an arbitrarily shaped coherent
source.  We then split the source into a collection of lasers at
slightly different wavelenths, that is, make it a narrow-band
incoherent source.  We illustrate with simulations how this results in
a mess of transient interference patterns, which are unobservable
because there are not enough photons.  We then introduce a property
which, like the fringe spacing in two-slit interference, is
well-behaved --- namely the intensity correlation.  We also briefly
discuss the differences with respect to Michelson-type interferometry.
In Section~\ref{sec:sn} we discuss how the time resolution of photon
detectors enters into the practically observable signal and noise.  We
note the inverse problem of reconstructing the source from the
observables, but do not discuss inversion algorithms.  Another topic
we will not attempt to address is speckle imaging.  Like intensity
interferometry, speckle imaging is also undergoing a revival of
interest \citep[for some recent developments,
  see][]{2018PhT....71k..78H} and that it is somehow related to
intensity interferometry has been known from the earliest days
\citep{1970A&A.....6...85L} and a common theory explaining and
contrasting both is desirable.

\begin{figure}
\includegraphics[height=.3\vsize]{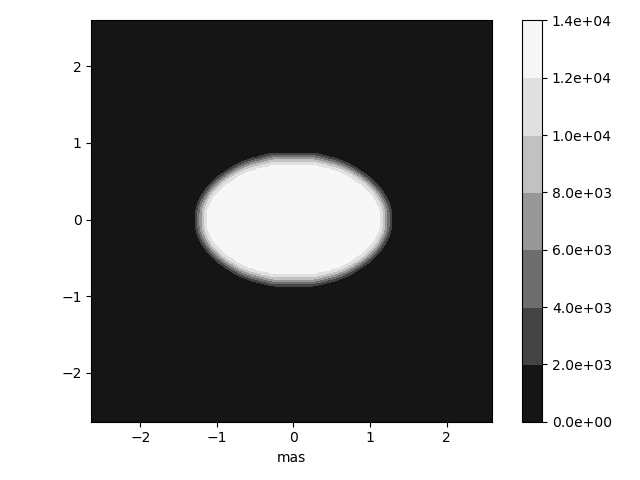}\\
\includegraphics[height=.3\vsize]{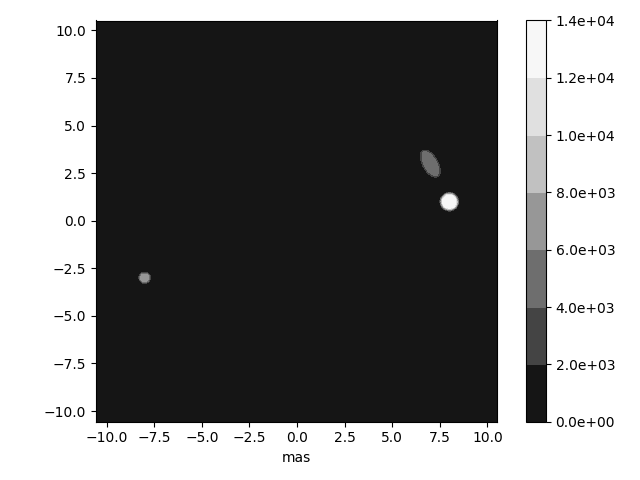}\\
\includegraphics[height=.3\vsize]{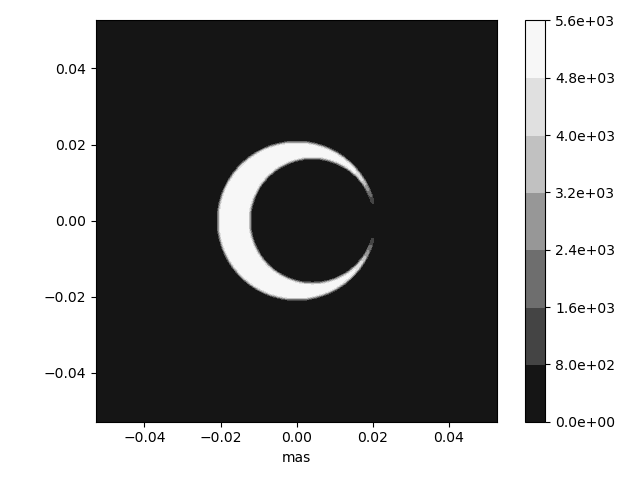}%
\caption{Example sources with effective temperature in K. See the last
  part of Section~\ref{sec:intro} for details.\label{fig:src}}
\end{figure}

There is a wealth of possible targets for next-generation
interferometers \citep[see Figure~10 of][]{2013APh....43..331D} but as
illustrative examples we will consider three simple models consisting
of blackbody sources in simple shapes. There are shown in
Figure~\ref{fig:src}.
\begin{itemize}
\item First and simplest we have an ellipse of $\rm 2.4\,mas\times
  1.6\,mas$ at $T_{\rm eff}=12500\rm\,K$, which approximates Achernar.
  This star was actually observed by \cite{1974MNRAS.167..121H} but
  without allowing for its rotational flattening, and hence measuring
  only a mean angular diameter.  The flattening was measured
  by \cite{2003A&A...407L..47D} using Michelson-type interferometry.
\item Next we have a three-component source approximating the Algol
  system, also based on reconstructions from Michelson-type
  interferometry \citep{2010ApJ...715L..44Z,2012ApJ...752...20B}.
\item Then we have a crescent source at $T_{\rm eff}=5000\rm\,K$,
  which does not represent any particular object but is reminiscent of
  two kinds of interesting systems.  The angular size and pattern
  roughly mimics the M87 black hole \citep[cf.][]{2013MNRAS.434..765K}
  though the brightness at optical wavelengths is totally speculative.
  Alternatively, we can imagine a subdwarf star some tens of pc away,
  with a giant planet in transit.
\end{itemize}
These systems are shown in Figure~\ref{fig:src}.  We choose the
frequency as the peak of the visible range ($\nu=540\rm\,THz$ or
$\lambda=555\rm\,nm$).  In the course of this paper we will compute
the expected signal and noise for each of these systems.

\section{Transient interference}\label{sec:transient}

In optics, coherent light produces interference, whereas incoherent
light simply adds.  Coherent vs incoherent is, however, not a
dichotomy.  Incoherent light filtered to a narrow band becomes
partially coherent, and produces transient interference effects, which
are only detectable with ultra-fast intensity measurements.  This fact
is the heart of the Hanbury Brown and Twiss effect, and the basis of
intensity interferometry, but conventionally it is stated in terms of
correlations and bunching.  Here we will show that HBT can be
considered as an extension of classical diffraction.

We will assume that (i)~ordinary light is equivalent to a
superposition of many lasers over a range of frequencies, and (ii)~the
classical intensity, suitably normalised, corresponds to the
probability of detecting photons.

\subsection{Brightness}

Consider a channel in which light is filtered to a narrow band of
width $\Delta\nu$ around a frequency $\nu$, and to a single
polarisation state.  There can be an independent channel for the
orthogonal polarisation, and further channels at other frequencies,
but let us consider this one channel.  In this channel, let
$|S(\bO)|^2\,\Delta\nu$ be the photon flux coming from direction $\bO$
on the sky.  $|S(\bO)|^2$ will have units of photons~$\rm
m^{-2}\,sr^{-1}\,s^{-1}\,Hz^{-1}$.  In particular
\begin{equation}\label{eq:teff}
|S(\bO)|^2 = \frac{\nu^2/c^2}{e^{h\nu/(kT(\bO))}-1}
\end{equation}
corresponds to a blackbody source with a temperature map of $T(\bO)$.
Even if the source is not blackbody, Eq.~\eqref{eq:teff} can still be
used within a narrow band to give the photon flux in terms of an
effective temperature.

Integrating over the source, let us write
\begin{equation}
\pcoh = \int |S(\bO)|^2 \, d^2\bO 
\end{equation}
for the photons~$\rm m^{-2}\,s^{-1}\,Hz^{-1}$ coming from the source.
The quantity $\Phi$ is a spectral photon flux, and is sometimes called
the count degeneracy \citep[see, e.g.,][]{Goodman2015}.

Brightness in janskys or optical magnitudes can be replaced by $\Phi$.
To do so, note first that the energy flux in a band (including both
polarisations) will be $2h\nu\,\Delta\nu\,\pcoh$.  Hence $2h\nu\,
\pcoh$ will be the spectral flux density in $\rm
W\,m^{-2}\,Hz^{-1}$. (Recall that $1\rm\,Jy$ is $10^{26}$ of this
unit.)  In terms of AB magnitudes we have
\begin{equation}\label{eq:ABmag}
2h\nu \, \pcoh = 10^{-22.44-{\rm AB}/2.5} \rm \;W\,m^{-2}\,Hz^{-1}
\end{equation}
This can be rearranged and rounded to the convenient expression
\begin{equation}
\pcoh \approx 10^{-4-{\rm AB}/2.5} \frac\lambda{1\mu\rm m} {\rm\; m^{-2}}
\end{equation}
The three example systems in Figure~\ref{fig:src} have
\begin{equation}
\pcoh = \left\{
\begin{matrix}
3.1\times 10^{-5} \\
6.0\times 10^{-6} \\
1.8\times 10^{-10} \\
\end{matrix}
\right\} \hbox{photons}\ \rm m^{-2}\,s^{-1}\,Hz^{-1}
\end{equation}
respectively.

The product $A\,\pcoh$, where $A$ is an effective detector area (that
is, light-collecting area times system throughput and detector
efficiency), will be important later.

\subsection{Complex amplitude}

If the source were coherent and had a monochromatic frequency
$\nu$, we could write
\begin{equation} \label{eq:amplSc}
S(\bO) \, e^{2\pi i\,\nu t}
\end{equation}
for the probability amplitude of a photon to come from direction
$\bO$.  The probability amplitude at a location $\bb$ on the ground
could be written as
\begin{equation} \label{eq:amplVc}
\pha(\bb) \, e^{2\pi i\,\nu t}
\end{equation}
which corresponds to a brightness distribution $|\pha(\bb)|^2$ on the
ground.  We will assume that $\bO$ is always a small solid angle, and
further that it is near the zenith.  For sources not at zenith one
needs to apply a standard rotation (given in
Appendix~\ref{sec:matrices}) to $\bb$.

The two amplitudes will be related by the classical Fraunhofer
diffraction formula
\begin{equation}\label{eq:fraunhofer}
\pha(\bb) \propto \int e^{2\pi i\,(\nu/c)\,\bb\cdot\bO} \, S(\bO) \, d^2\bO
\end{equation}
with the proportionality constant chosen to satisfy
\begin{equation}\label{eq:pcoh}
\left\langle \, |\pha(\bb)|^2 \right\rangle = \pcoh
\end{equation}
The proportionality constant is given in Eq.~\eqref{eq:fraunhofer2} in
Appendix~\ref{sec:numer}.

\subsection{Narrowband light}

Let us now consider an incoherent source filtered to a narrow band
around $\nu$, giving a frequency spectrum $W(\nu')$.  We have already
used the bandwidth $\Delta\nu$.  Let us now define it more precisely
as
\begin{equation}\label{eq:bwidth}
\Delta\nu = \int\! W(\nu') \, d\nu' = \int\! W^2(\nu') \, d\nu'
\end{equation}
which also fixes the normalisation of the bandpass $W(\nu')$.  Next,
let us introduce
\begin{equation}\label{eq:cohf}
\gamma(t) = \frac1{\Delta\nu} \int e^{2\pi i\,(\nu'-\nu)t}\, W(\nu')\, d\nu'
\end{equation}
which is a shifted Fourier transform of $W(\nu')$.  We then
define the time scale
\begin{equation}\label{eq:tcoh}
\Delta\tau = \int\! |\gamma(t)|^2 \, dt
\end{equation}
which we will call the coherence time.\footnote{The precise definition
  of $\Delta\nu$ and $\Delta\tau$ varies in the literature.  We are
  following Eqs.~(5.28--5.30) of \cite{1965RvMP...37..231M}, with our
  $W$ being $\Delta\nu\times w$ in that work.}
As we will see below, the coherence time is an indicator of how long
Fraunhofer diffraction patterns persist.
The Parseval relation corresponding to Eq.~\eqref{eq:cohf} implies
\begin{equation}
\Delta\tau = (\Delta\nu)^{-1}
\end{equation}
To take an example, suppose we have a narrowband filter of
$\Delta\lambda=1\rm\,nm$ around $\lambda=555\rm\,nm$.  The coherence
time will then be
$\Delta\tau=\lambda^2/(c\Delta\lambda)\approx1\rm\,ps$.

For narrowband light, the complex amplitude \eqref{eq:amplSc} on the
sky will be replaced by
\begin{equation}
|S(\bO)| \; e^{i\rph(\bO)} \times \frac1{\Delta\nu}
\int e^{2\pi i\,\nu't}\, W(\nu')\, d\nu'
\end{equation}
where $\rph(\bO)$ is an initial phase, which is randomly different
across the source.  The ground amplitude \eqref{eq:amplVc} is
proportional to
\begin{equation}
\begin{aligned}
|S(\bO)| \; e^{i\rph(\bO)} \times \frac1{\Delta\nu}
\int & e^{2\pi i(\nu'/c)\,\bb\cdot\bO} \\
     & e^{2\pi i\,\nu' t}\, W(\nu') \; d\nu' \,d^2\bO
\end{aligned}
\end{equation}
We assume $|\nu'-\nu|\ll\nu$ and replace
$$
e^{2\pi i(\nu'/c)\,\bb\cdot\bO} \rightarrow
e^{2\pi i(\nu/c)\,\bb\cdot\bO}
$$
because there is no multiplication by $t$.  Then the sky amplitude can
be written as
\begin{equation}
S(\bO,t) \, e^{2\pi i\, \nu t}
\end{equation}
where
\begin{equation}\label{eq:narrowband}
S(\bO,t) = |S(\bO)| \; e^{i\rph(\bO)} \times \gamma(t)
\end{equation}
and the ground amplitude can be written as
\begin{equation}
\pha(\bb,t) \, e^{2\pi i\, \nu t}
\end{equation}
where
\begin{equation} \label{eq:amplV}
\pha(\bb,t) \propto \int e^{2\pi i\,(\nu/c)\,\bb\cdot\bO} \,S(\bO,t)\, d^2\bO
\end{equation}
Eqs.~\eqref{eq:narrowband} and \eqref{eq:amplV} describe Fraunhofer
diffraction with a slow time dependence.  We see that a small spread
in frequency around $\nu$ produces a slow variation in the brightness
distribution $|\pha(\bb,t)|^2$ on the ground.  The larger the
frequency spread, the faster the brightness will vary.

The photon flux on the ground is given by
\begin{equation}
|\pha(\bb,t)|^2\;\Delta\nu \quad\hbox{or}\quad |\pha(\bb,t)|^2 \; (\Delta\tau)^{-1}
\end{equation}
and $|\pha(\bb,t)|^2$ nominally has units of photons $\rm
s^{-1}\,m^{-2}\,Hz^{-1}$.  It is more useful, however, to think of
$|\pha(\bb,t)|^2$ as photons $\rm m^{-2}$ per coherence time.
Widening the bandpass will increase the photon flux, but the number of
photons per unit area within a coherence time will remain the same.
This fact will be important when we consider signal and noise.

\begin{figure*}
\includegraphics[width=.47\hsize]{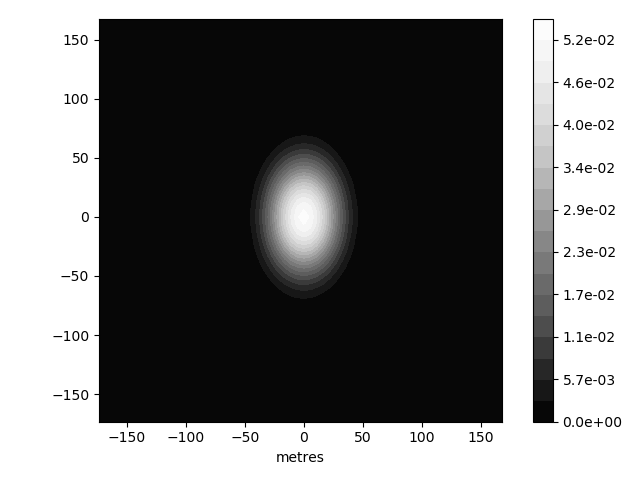}
\includegraphics[width=.47\hsize]{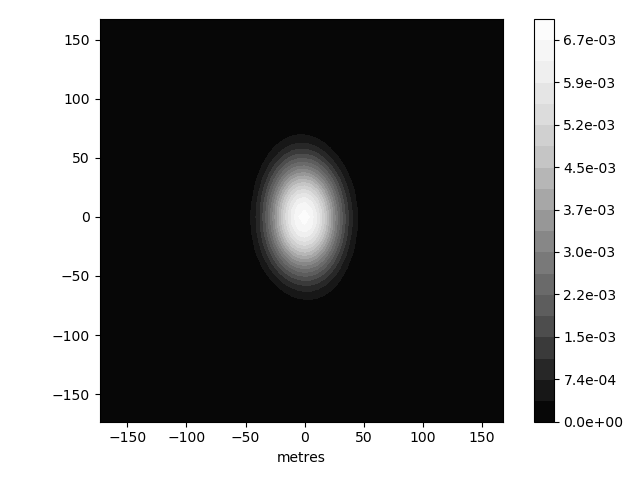}\\
\includegraphics[width=.47\hsize]{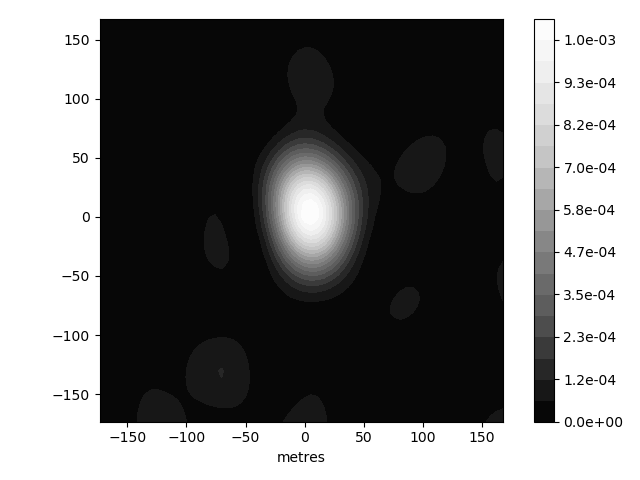}
\includegraphics[width=.47\hsize]{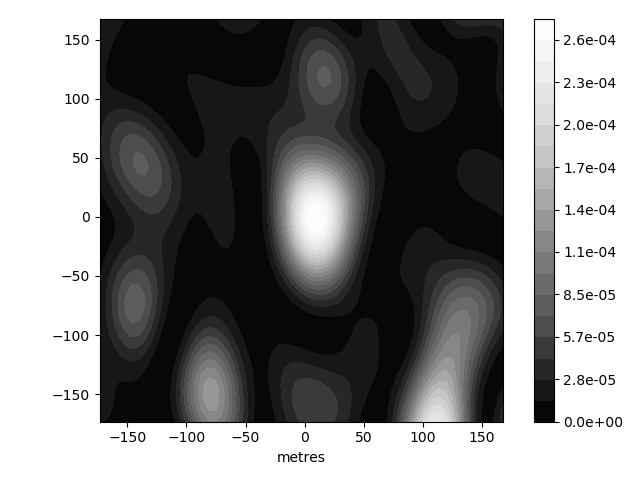}\\
\includegraphics[width=.47\hsize]{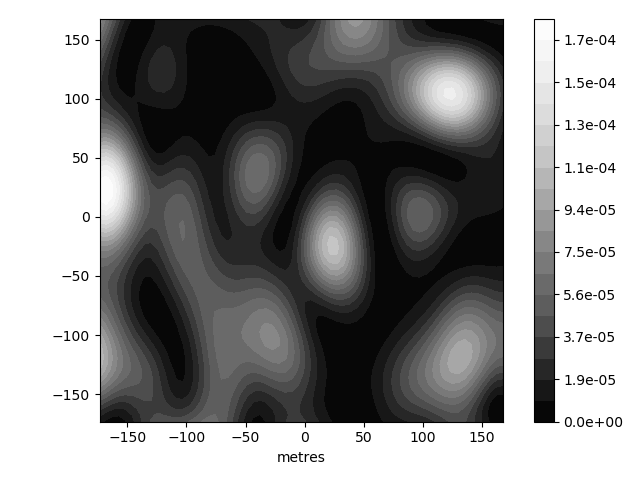}
\includegraphics[width=.47\hsize]{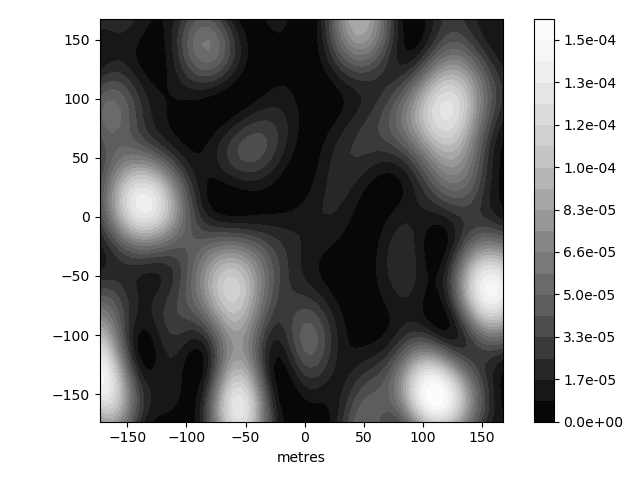}
\caption{Simulation of decoherence of an initially coherent
  interference pattern.  Shown is the ground brightness
  $|\pha(\bb,t)|^2$ in photons $\rm m^{-1}$ per coherence time for the
  Achernar-like example source.  Panels (in reading order, from top
  right to bottom left) are at $t=\Delta\tau,\dots,5\Delta\tau$.  The
  frequency band is Lorentzian, and initial peak fades exponentially,
  as expected from Eq.~\eqref{eq:fading}.  \label{fig:devel}}
\end{figure*}

\subsection{Decoherence}

As an example of a spectral bandpass, consider
\begin{equation}\label{eq:freqfilter}
W(\nu') = \frac2{1+(2\pi\,\Delta\tau(\nu'-\nu))^2}
\end{equation}
which one can verify satisfies the normalisation \eqref{eq:bwidth}.
Physically, it is a Lorentzian spectrum with an arbitrary phase factor
depending on $\bO$.  For this spectrum we can then work out
\begin{equation}\label{eq:fading}
S(\bO,t) = |S(\bO)|
           \exp\left(-\left|\frac t{\Delta\tau}
                     -\frac{\rph(\bO)}{2\pi}\right|\right)
\end{equation}
which describes a flickering source: pulses rise at different $\bO$ at
random times, and then fade exponentially, while the diffraction
pattern on the ground flickers accordingly.  Integrating, we can verify
\begin{equation}
\int |S(\bO,t)|^2 \, dt = \Delta\tau \, |S(\bO)|^2
\end{equation}
as implied by Eq.~\eqref{eq:narrowband} above.

We can think of the source as the sum of many lasers with slightly
different frequencies, the frequencies being distributed according to
the Lorentzian~\eqref{eq:freqfilter}.  For simulations we set the
phase of all frequency components to 0 at $t=0$.  Thus the source is
coherent at $t=0$, but as time passes the frequency differences make
the source incoherent.

Figure~\ref{fig:devel} illustrates this decoherence.  The figure shows
$|\pha(\bb,t)|^2$ from the Achernar-like source at
$t=0,\Delta\tau,\ldots,5\Delta\tau$.  At $t=0$ there is a single
diffraction peak.  The peak remains till $t=3\Delta\tau$ but its
intensity fades as $e^{-2t/\Delta\tau}$ as expected from
Eq.~\eqref{eq:fading}.  By the last panel $t=5\Delta\tau$ the original
diffraction peak is gone but many new peaks have appeared.  This panel
is typical of $|\pha(\bb,t)|^2$ for incoherent light.  In real life, we
are never close to a coherent epoch for astronomical sources, so the
scenario in this figure is completely artificial.  For the following
figures we will consider only random, hence incoherent, epochs.
Setting $t\gg\Delta\tau$ in Eq.~\eqref{eq:photonexp} below ensures
that.

\begin{figure}
\includegraphics[height=.3\vsize]{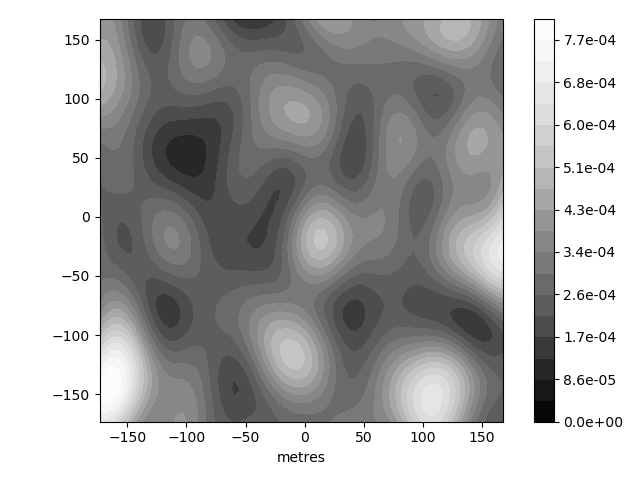}\\
\includegraphics[height=.3\vsize]{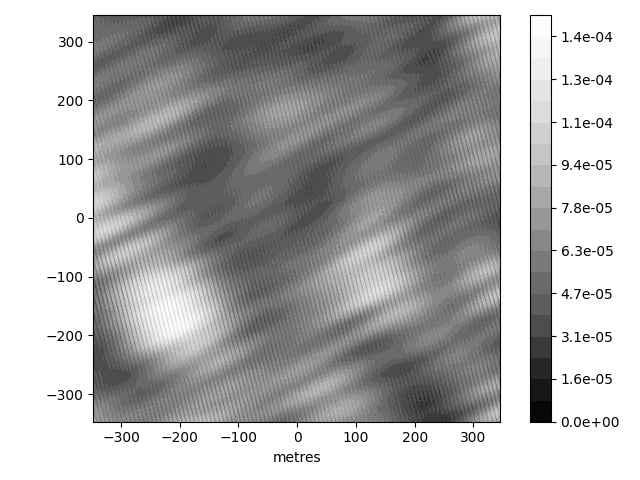}\\
\includegraphics[height=.3\vsize]{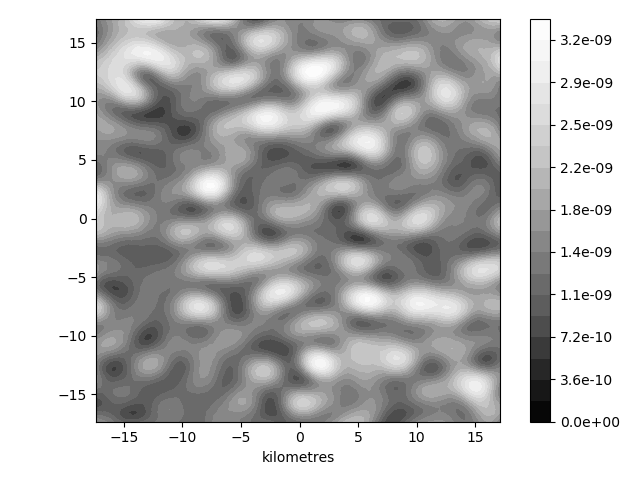}%
\caption{Transient interference patterns $X(\bb,\Delta
  t=10\Delta\tau)$ in photons $\rm m^{-2}$ (see
  Eq.~\ref{eq:photonexp}) for the three example sources from
  Figure~\ref{fig:src}. \label{fig:intens}}
\end{figure}

\subsection{Transient intensity}

Although the intensity pattern on the ground changes completely within
$5\Delta\tau$, it does not average out to uniform within that time.
Consider now the expected photons~$\rm m^{-2}$ over a time slice
$\Delta t$ following time $t$.  Call this $X(\bb,\Delta t)$ for photon
exposure (there is no standard name).  We have
\begin{equation}\label{eq:photonexp}
X(\bb,\Delta t) = \Delta\nu \int_t^{t+\Delta t}
|\pha(\bb,t')|^2 \, dt'
\end{equation}
with $\Delta\nu=1/\Delta\tau$ as before.  The uncertainty principle
requires
\begin{equation}
\Delta t > \Delta\tau \quad \hbox{(typically $\Delta
  t\gg\Delta\tau$)}.
\end{equation}
Figure~\ref{fig:intens} shows $X(\bb,\Delta t=10\Delta\tau)$ for each
of our example sources.  (The three panels shown are 8-fold, 16-fold,
and 4-fold zooms of the full simulations.)  As can be read off from
Figure~\ref{fig:intens}, a bright blob is expected to yield less than
one photon over a time interval $10\Delta\tau$.  Integrating for many
$\Delta\tau$ longer will not help trace the blobs, because the pattern
keeps changing.  The patterns in the simulation are thus too faint to
be photographed.

Although unobservable with ordinary light, patterns similar to
Figure~\ref{fig:intens} can be observed in laboratory experiments with
pseudo-thermal (or quasi-thermal) light, which is light with an
adjustable coherence time, produced say by moving ground glass across
a coherent source.  Some early but very nice examples appear in
\cite{1964AmJPh..32..919M}.

\subsection{Correlation}

The transient patterns in Figure~\ref{fig:intens} clearly do contain
information about the source map $|S(\bb)|^2$ but in a highly
distorted form.  We now see how to extract that information.

Consider the product $\pha(\bb_1,t) \, \pha^*(\bb_2,t)$ which
represents the instantaneous correlation between the amplitudes at two
points on the ground.  Note that complex amplitudes averages to zero,
hence there is no constant term to subtract in the correlation.  From
the Fraunhofer diffraction formula \eqref{eq:amplV} we see that the
instantaneous correlation will be an integral over $\bO$ and $\bO'$
with the integrand containing factors of $S(\bO,t)\,S^*(\bO',t)$.
Because the light is chaotic, over long times the latter factors tend
to cancel if $\bO\neq\bO'$, and only $|S(\bO,t)|^2$ contributes.  This
fact is essentially the van~Cittert-Zernike theorem, which has many
interesting consequences \citep[and not only for light waves ---
  see][]{2010PhT....63c..11K}.  The consequence of interest here is
that the time-averaged correlation simplifies to
\begin{equation} \label{eq:vCZ}
  \begin{aligned}
    \langle \, \pha(\bb_1,t) \, \pha^*(\bb_2,t) \rangle \propto
    \int & e^{2\pi i\,(\nu/c)\,(\bb_1-\bb_2)\cdot\bO} \\
         & \left\langle \, |S(\bO,t)|^2 \right\rangle \, d^2\bO
\end{aligned}
\end{equation}

Let us write the time average of the right-hand side of \eqref{eq:vCZ}
as $\vis(\bb_1-\bb_2)$, with the normalisation $\vis(0)=1$.  That is,
$\vis(\bb)$ is the spatial Fourier transform of the source intensity
(not the source amplitude), with any source flickering time-averaged
out.  The time average of Eq.~\eqref{eq:vCZ} is then
\begin{equation} \label{eq:visib}
\frac{\langle \pha_1 \, \pha_2^* \rangle}
{\sqrt{\left\langle \, |\pha_1|^2 \right\rangle \,
       \left\langle \, |\pha_2|^2 \right\rangle}}
= \vis_{12}
\end{equation}
using the abbreviated notation $\pha_1 \equiv \pha(\bb_1,t)$,
$\pha_2 \equiv \pha(\bb_2,t), \vis_{12}\equiv\vis(\bb_1-\bb_2)$.  The
physical meaning of this equation is that the time-averaged amplitude
correlation on the ground is given by the normalised spatial Fourier
transform of the source brightness.

The quantity $\vis(\bb)$ is known as the complex visibility, and
interferometry is all about measuring as much as possible of
$\vis(\bb)$.  The difficulty is that measuring $\pha(\bb_1,t) \,
\pha^*(\bb_2,t)$ requires controlling the optical path difference
between $\bb_1$ and $\bb_2$ to sub-wavelength precision.

Intensity interferometry adopts a different strategy.  First we note,
that as we have seen from Figures~\ref{fig:devel} and
\ref{fig:intens}, $|\pha(\bb,t)|^2$ varies on the comparatively slow
time scale of $\Delta\tau$.  This makes the time average $\left\langle
\, |\pha_1|^2 \, |\pha_2|^2 \right\rangle$ measurable without
requiring sub-wavelength precision.  We are then helped by an identity
variously known as Isserlis' theorem and Wick's theorem.  Of interest
here is the case expressing $\langle \pha_1 \ldots \pha_N \, \pha_N^*
\ldots \pha_1^* \rangle$ in terms of products $\langle \pha_m \,
\pha_n^* \rangle$ \citep[see e.g., Eq.~10.27
  in][]{1963PhRv..131.2766G}.  The simplest instance is
\begin{equation}\label{eq:glauber}
\langle \pha_1 \, \pha_1^* \, \pha_2 \, \pha_2^* \rangle =
\langle \, |\pha_1|^2 \rangle \langle \, |\pha_2|^2 \rangle +
\left| \langle \pha_1 \, \pha_2^* \rangle \right|^2
\end{equation}
relating the intensity at two points to the visibility.  Rearranging,
we have
\begin{equation}\label{eq:correl}
\frac{\left\langle \, |\pha_1|^2 \, |\pha_2|^2 \right\rangle}
     {\left\langle \, |\pha_1|^2 \right\rangle
      \left\langle \, |\pha_2|^2 \right\rangle} - 1
= |\vis_{12}|^2
\end{equation}
As we see, correlating intensity at two points loses the phase
information in $\vis_{12}$.  The transient interference pattern
contains more information, which could in principle be recovered by
higher-order correlation.  In particular, three-point intensity
correlation includes a dependence on
\begin{equation}
\mathop{\rm Re}
\big[\,
     \vis(\bb_1-\bb_2)\;\vis(\bb_2-\bb_3)\;\vis(\bb_3-\bb_1)
\,\big]
\end{equation}
\citep[see e.g.,][]{2015MNRAS.446.2065W}.  Still higher orders also
possible \citep{1983JAP....54..473F,2014MNRAS.437..798M}.

\subsection{Summary}

In this section, we have considered three forms of optical
interference.  Let us summarise these.

First we have Fraunhofer diffraction, which we write concisely as
follows.
\def\upsp{\vrule width 0pt height 15pt} \def\calF{{\cal F}}
\begin{equation}
\begin{aligned}
\pha &= \calF(S) \\
|\pha|^2 &= \calF(S\circ S^*) \upsp \\
\end{aligned}
\end{equation}
Here $\calF$ denotes a Fourier transform, and expressions with $\circ$
are auto-correlations.  The complex amplitude $\pha(\bb)$ on the
ground is the Fourier transform of the complex amplitude $S(\bO)$ of
the source in the sky.  If sources were coherent (if phases on
different parts of the source were correlated) we would get
interference fringes on the ground.  But phases on the source are not
correlated, hence $S\circ S^*$ is chaotic.  The corresponding
intensity $|\pha|^2$ on the ground is also chaotic, and time-averages
to uniform illumination.  This is illustrated in
Figures~\ref{fig:devel} and \ref{fig:intens}.

Astronomical Michelson interferometry makes use of the
van~Cittert-Zernike theorem
\begin{equation}\label{eq:vCZ2}
\pha\circ\pha^* = \calF\left(\,|S|^2\right) \\
\end{equation}
which as written here can be considered a corollary of the Fraunhofer
diffraction integral.  The source can be incoherent, and there are no
interference fringes in $|\pha|^2$.  Instead, fringes in
$\pha\circ\pha^*$ are observable if phase coherence can be maintained
across the telescope.

Finally we have intensity interferometry, which actually works because
the source is not coherent, and the phase in $\pha$ is chaotic.
Complex fields with random phases have the property that
\begin{equation}
|\pha|^2 \circ |\pha|^2 = |\pha|^4 + |\pha\circ\pha^*|^2
\end{equation}
and by exploiting this, $|\pha\circ\pha^*|^2$ can be inferred from
intensity correlation without the need for phase coherence.  The
intensity patterns in Figure~\ref{fig:intens} are chaotic, but their
auto-correlations are well-behaved.  We will see this below in
Figure~\ref{fig:abstract}.

In the next section, we will be especially concerned with $|\vis|$ and
$\pcoh$.  These are nothing but the normalised form of
$|\pha\circ\pha^*|$ and the normalisation factor respectively (see
Eqs.~\ref{eq:visib} and \ref{eq:pcoh}).

\section{Signal and noise}\label{sec:sn}

The apparatus in intensity interferometry is a light bucket, which
counts photons over an effective area $A$ on the ground over some
time-slice $\Delta t$. Let us write $N(\bb)$ for the number of photons
detected in a time-slice $\Delta t$.  It will be usually 0, sometimes
1, but its expectation value will be
\begin{equation}
E\big(N(\bb)\big) = A \, X(\bb,\Delta t)
\end{equation}
The basic observable in intensity interferometry is the time-averaged
correlation between $N(\bb_1)$ and $N(\bb_2)$.

\subsection{Photon correlation}

Using the abbreviations $N_1$ and $N_2$
\begin{equation}\label{eq:gdef}
\gobs_{12} \equiv \frac{\langle N_1 N_2 \rangle}
                       {\langle N_1 \rangle \langle N_2 \rangle} - 1
\end{equation}
is the observable HBT correlation.  This definition is a modified
version of the formula~\eqref{eq:correl} which gives $|\vis_{12}|^2$.
A modification is required because in practice $\Delta t \gg
\Delta\tau$, and HBT-correlated counts only occur within
$\sim\Delta\tau$ of each other (before the transient interference
pattern changes).  Let us divide a time interval $\Delta t$ into
$\Delta t/\Delta\tau$ time-slices of duration $\Delta\tau$ each, and
approximate the light as being perfectly coherent within a time slice
and completely incoherent between different time slices.  The expected
number of counts in a coherence time is $A\,\pcoh$.  The number of HBT
correlated counts in $\Delta t$ will be
\begin{equation}\label{eq:sig}
\langle N_1 N_2 \rangle - \langle N_1 \rangle \langle N_2 \rangle
\approx (\Delta t/\Delta\tau) \; (A\,\pcoh)^2 \; |\vis_{12}|^2
\end{equation}
whereas the total correlated counts will be
\begin{equation}\label{eq:n2}
  \langle N_1 \rangle \langle N_2 \rangle =
  (\Delta t/\Delta\tau)^2 \; (A\,\pcoh)^2
\end{equation}
Together these give
\begin{equation}\label{eq:signal}
\gobs_{12} \approx \frac{\Delta\tau}{\Delta t} \, |\vis_{12}|^2
\end{equation}
which is the observable correlation in intensity interferometry.
Measuring the correlation with detectors close together will give
$\gobs(0)$ which is $\approx\Delta\tau/\Delta t$ since $\vis(0)=1$ by
definition.  The precise proportionality factor here will depend on
the details of the frequency bandpass and time response of the photon
detectors.

The instruments in first-generation intensity interferometry did not
count photons explicitly.  Instead, there were currents proportional
to the photon counts, and the currents were correlated in
hardware.\footnote{The time resolution $\Delta t$ also did not appear
  explicitly, but as the reciprocal of twice the frequency bandwidth
  (known as the ``electronic bandwidth'') in the correlating
  hardware.}  In the definition \eqref{eq:gdef} of $\gobs_{12}$ it
does not matter whether we use counts or intensities.  Present-day
technology, on the other hand, uses digitised signals from individual
photon detections.  This makes is interesting to consider variants of
$g_{12}$ that are meaningful for photons counts but not for
intensities.  In the following sections, we consider two such
variants.

\subsection{A scaled correlation}

Consider the quantity
\begin{equation}\label{eq:snr}
\SNR_{12} \equiv
     \frac{\langle N_1 N_2 \rangle - \langle N_1 \rangle \langle N_2 \rangle}
          {\sqrt{\langle N_1 \rangle \langle N_2 \rangle}}
\end{equation}
which is the correlation times the geometric mean of counts in a time
slice.  Referring back to Eqs.~\eqref{eq:n2} and \eqref{eq:signal} and
noting that the average number of counts in a detector is $A\,\Phi$,
we have
\begin{equation}\label{eq:snrsimple}
\SNR(\bb) \approx A \, \pcoh \, |\vis(\bb)|^2
\end{equation}
where $A$ is understood as the geometric mean of the effective areas.
Comparing with Eqs.~\eqref{eq:signal} and \eqref{eq:n2} we can see
that $\SNR_{12}$ has the interpretation of signal-to-noise (SNR) per
data point.  Of course, $\SNR_{12}$ is really an upper limit on the
achievable SNR.  The actual SNR will be lower, because of additional
noise sources.  We will call $h(\bb)$ the scaled correlation.

It may be that one is interested in the difference of two cases, such
as the on-transit and off-transit epochs of an exoplanetary system
\cite[cf. Fig.~11 of][]{2014SPIE.9146E..0ZD}.  For such situations, we
define a differential signal which still has the interpretation of SNR
per data point.  Let us define
\begin{equation}\label{eq:snrdiff}
\sigma_A \equiv \SNR_A(0)
\end{equation}
which is approximately the numerator of Eq.~\eqref{eq:snr} and hence
can be interpreted as the noise level.  Analogously, $\sigma\SNR(\bb)$
can be interpreted as the signal.  Then the difference in signal
between two cases, scaled by the combined noise, will be
\begin{equation}\label{eq:dsnr}
\Delta\SNR = \frac{\sigma_A\SNR_A - \sigma_B\SNR_B}
                  {\sqrt{\sigma_B^2+ \sigma_B^2}}
\end{equation}
where the subscripts refer to the two cases.

\subsection{Properties of the scaled correlation}

The scaled correlation $\SNR(\bb)$ is a small number which gets
measured once per time-slice $\Delta t$, with a noise of unity from
photon statistics.  Over a long observation, the noise falls to
$(\Delta t/t_{\rm obs})^{1/2}$.  The NSII used $\Delta t=\rm10\,ns$.
Nowadays, $\Delta t=\rm1\,ns$ is common and $\Delta t= \rm0.1\,ns$ is
possible \citep{2017MNRAS.467.3048P}.  Thus, one night of observing
could have $>10^{14}$ data points.  The vast majority of these data
points may have zero photons detected, but with so many data points
$\SNR\sim10^{-6}$ would be measurable.  Once $\SNR(\bb)$ is measured,
$\vis(\bb)$ will also be determined, as the two differ only in
normalisation.

Interestingly, the bandwidth (and hence the coherence time) does
not appear in the scaled correlation.  This is because decreasing the
bandwidth decreases the photon count, but correspondingly increases
the coherence time, so $A\pcoh$ remains the same.  With SNR not
depending on bandwidth, it is possible to split a narrow spectral band
into multiple narrower bands, each having the same SNR as before.  (It
is understood that the coherence time $\Delta\tau$ is shorter than the
detector time resolution $\Delta t$.  Narrowing the bandpass will help
only till $\Delta\tau\approx\Delta t$.)  Some experimental detectors
\citep[see][]{2016SPIE.9907E..1WH} have multiple simultaneous channels
observing in separate narrow spectral bands.  In a laboratory setting
kilo-pixel photon detection with $\Delta t<1\rm\,ns$ has been achieved
\citep{2019OExpr..2735279W}.  If the latter are some day installed on
intensity interferometers, an observing night would yield
$\sim10^{17}$ data points, and then even $\SNR=10^{-8}$ would be
viable.

In $\pcoh$ we have considered only one polarisation channel.  Detector
designs so far do not separate the polarisation channels.  When this
is done, the two polarisation states behave like two unseparated
spectral channels: both signal and noise double, leaving the SNR the
same.  If photons in two polarisation channels are correlated
separately, that should give two simultaneous measurements with the
same SNR.

That $\SNR$ is independent of both $\Delta t$ and $\Delta\tau$ is
something peculiar to two-point HBT.  For $N$-point correlations
the scaled correlation is
\begin{equation}
\SNR^{(N)} \approx \vis^{(N)} \times (A\pcoh)^{N/2}
\left(\frac{\Delta\tau}{\Delta t}\right)^{N/2-1}
\end{equation}
\citep[cf.][]{2014MNRAS.437..798M}.  We will not discuss higher-order
HBT in this paper, but just remark here that it is much more difficult
to observe, because $\pcoh$ is raised to a higher power in the SNR.

\subsection{A correlation density}

Another possible variant of the correlation is
\begin{equation}\label{eq:sigf}
\sigf(\bb_2-\bb_1) \equiv \frac{\langle X(\bb_1) X(\bb_2) \rangle}
                               {\langle X \rangle}
                   - {\langle X \rangle}
\end{equation}
having dimensions of inverse area.  We may call the correlation
density.  Clearly $\sigf(\bb)=\langle X\rangle\,\gobs(\bb)$ and hence
\begin{equation}\label{eq:sigfsimple}
\sigf(\bb) \approx \pcoh \; |\vis(\bb)|^2
\end{equation}
which depends only on the source.  The correlation density is
convenient when we do not have a particular observational setup in
mind, and wish to assess what instrumentation would be needed to
resolve a given source.  Examples follow.

\subsection{Additional noise sources}

Beyond the essential statistics of photons, any practical intensity
interferometer will have additional sources of noise.  Here we just
mention three, which are considered in simulations by
\cite{2013MNRAS.430.3187R}.

First, there is extra light which produces noise without signal.
Mirrors that are not of optical quality have a roughness that produces
a large point spread function.  For Cherenkov telescopes, the point
spread functions are an arc-minute or more \citep[see
  e.g.,][]{2015SPIE.9603E..07T} and this naturally lets in extra light
from the night sky.  The SNR per data point will be
\begin{equation}
\frac{A\,\pcoh}{1 + \pcoh^x\!/\pcoh} \, |\vis(\bb)|^2
\end{equation}
where $\pcoh^x$ is the spectral photon flux of the extraneous light.
If $\pcoh^x\ll\pcoh$ the total observing time needed will be
$\propto\pcoh^{-2}$.  If $\pcoh^x\gg\pcoh$ the observing time becomes
$\propto\pcoh^{-4}$.  Hence, intensity interferometry would be
effectively infeasible for sources fainter than the night sky within
the point spread function.

Second, optical-path differences across the collecting mirror and
elsewhere in the optics may reduce the effective $\Delta t$.  This is
especially a problem if a solar-furnace design is used for the
collecting mirrors, because solar furnaces have no use for isochronous
optical paths.

Third, the response of the photon detectors may not be uniform.
\cite{1957RSPSA.242..300B} in their Eq.~(3.62) included a model for
non-uniform photo-electric response in their detectors, but current
detectors may need a different model.

\begin{figure}
\includegraphics[height=.3\vsize]{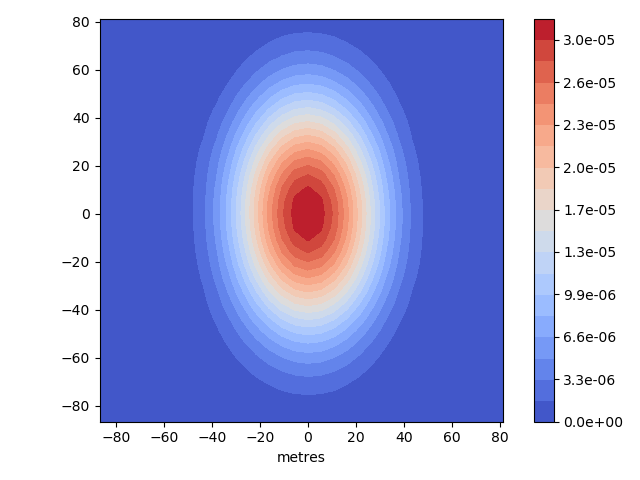}\\
\includegraphics[height=.3\vsize]{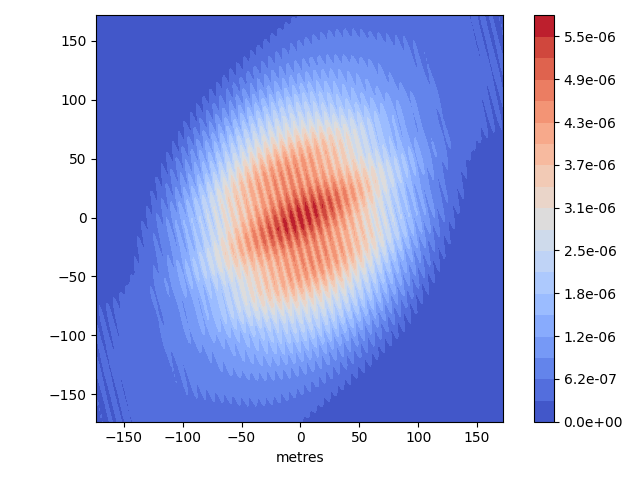}\\
\includegraphics[height=.3\vsize]{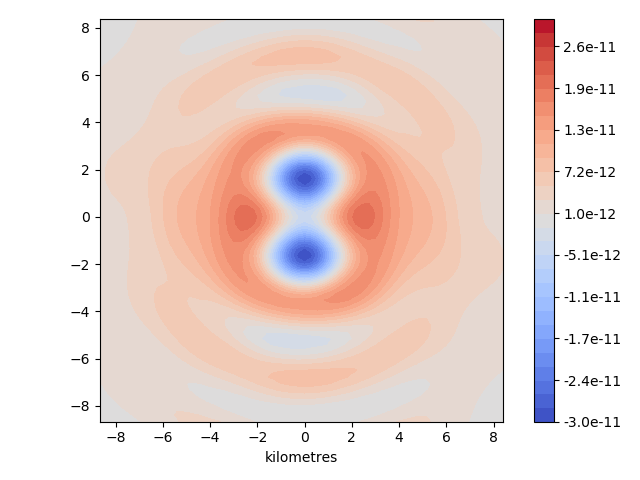}
\caption{Correlation density in coincidences $\rm m^{-2}$ per
  coherence time for the three example sources.  The upper and middle
  panels show $\sigf(\bb)$ from Eq.~\eqref{eq:sigf} while the lower
  panel shows $\Delta\sigf(\bb)$ from
  Eq.~\eqref{eq:dsigf}. \label{fig:abstract}}
\end{figure}

\subsection{The example systems}

Figure~\ref{fig:abstract} shows the simulated $\sigf(\bb)$ for our
three example sources, obtained from the simulated transient
interference patterns.  The simulated correlation $\gobs(\bb)$ is not
shown, since it only differs in the normalisation, but we have
verified that $\gobs(0)\simeq\Delta\tau/\Delta t$ in all cases.
Figure~\ref{fig:abstract} could also have been generated
Eq.~\eqref{eq:sigfsimple} without bothering with simulations of
$X(\bb)$, but it is nice to verify that simulating the transient
interference gives the expected result.

Note that the three panels in Figure~\ref{fig:abstract} are zoomed
twofold compared to Figure~\ref{fig:intens}.

For Achernar, we see that a collecting area of $0.1\rm\,m^2$ with a
throughput of order 50\% would bring $\SNR\sim10^{-6}$.  In the
optimistic scenario that this level of throughput is achieved, and
there are no other significant sources of noise, two amateur-grade
telescope with high-end photon counters and correlator would be enough
to measure the angular size and ellipticity of Achernar.

Algol represents a higher level of difficulty, both because it is
fainter and because there are three stars.  Here the adaptation of
Cherenkov telescopes for intensity interferometry currently under
development \citep[e.g.,][]{2019BAAS...51g.227K} are a promising
venue.  With mirror areas of $>10\rm\,m^2$ and baselines of up to
$100\rm\,m$, adequate SNR would be achievable.

For the crescent source, following the discussion around
Eq.~\eqref{eq:dsnr}, we show the differential quantity
\begin{equation}\label{eq:dsigf}
\Delta\sigf(\bb) = {\textstyle\frac1{\sqrt2}} (\sigf(\bb)-\bar\sigf(\bb))
\end{equation}
where $\bar\sigf(\bb)$ refers to a disc source with the same radius
and brightness.  For a transiting exoplanet, $\bar\sigf(\bb)$ would
correspond to an observable source, otherwise it is just a notional
reference.  Resolving the crescent source would not be possible in the
near future.  Cherenkov telescopes offer the prospect of
$10^3\rm\,m^2$ of collecting area and baselines of up to $2\rm\,km$.
The baseline is enough to resolve the crescent features, but the
collecting area would bring the SNR to only $\sim10^{-8}$ per data
point, which is not feasible to work with as yet. But brighter sources
of similar size are potential targets.

\subsection{The inverse problem}\label{subsec:inverse}

\begin{figure}
\includegraphics[height=.3\vsize]{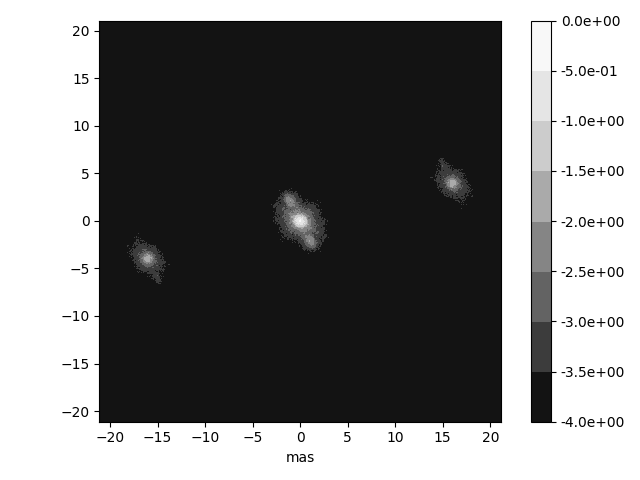}%
\caption{Dirty-map reconstruction of the Algol-like source (see
  subsection~\ref{subsec:inverse} for details). The scale is
  $\log_{10}$ with an arbitrary normalization. \label{fig:remap}}
\end{figure}

Reconstructing the source brightness distribution $|S(\bO)|^2$ from
incomplete information on $V(\bb)$ is an archetypical inverse problem.
In cases where there is a model for the source and only some parameter
fitting is required (say measuring angular size and ellipticity),
solving the inverse problem is not necessary.  But for general sources
the inverse problem will be important in intensity interferometry.
From Eq.~\eqref{eq:vCZ2} it follows that
\begin{equation}
|\pha\circ\pha^*|^2 = \calF\left(\,|S|^2\circ|S|^2\right) \,.
\end{equation}
To help understand this relation, in Figure~\ref{fig:remap} we have
taken $|\vis(\bb)|^2$ for the Algol-like source and plotted the
absolute value of its inverse Fourier transform.  The result, which as
we see has inversion symmetry, is not the original three-star source
brightness $|S(\bO)|^2$, but is clearly somehow related to it.  The
figure illustrates the unavoidable information loss in intensity
correlation.  In practice, one would not have a complete map of
$|\vis(\bb)|^2$, so taking the inverse Fourier transform of
$|\vis(\bb)|^2$ is not a feasible reconstruction method.  It is at best
comparable to a dirty-map reconstruction in radio interferometry.
However, algorithms for source reconstruction from intensity
correlation (including three-point correlation if available) have been
developed \citep{2015MNRAS.453.1999N}.

\section{Discussion}

This paper is about a conceptual question: what really is being
measured in intensity correlation?  We argue that intensity
interferometry can be thought of as measuring fringe sizes in a
transient interference pattern which cannot itself be observed,
because its bright fringes are sub-photon.  The essential idea has
always been implicit in the theory of HBT, just not made concrete,
because numerically simulating transient interference patterns was not
so easy when the theory was developed.

The notion of an interference pattern on the ground, even if
transient, suggests an interesting possibility for situations that
would currently be problematic.  As an example, consider the Algol
system, from which the smallest fringes are $<10\rm\,m$ apart in
visible light (see the middle panel in Figures~\ref{fig:intens}).
Mirrors up to 17~m in diameter are already in use for intensity
interferometry \citep{2020MNRAS.491.1540A} but too-large a light
bucket would average out the small fringes.  But suppose the light is
collected and brought to a focus by a large mirror, and then
re-collimated by a secondary mirror or lens.  The result would be a
miniature version of the interference pattern on the ground.  Photon
detection and correlation could be done on the miniaturised pattern,
with the help of further optical elements.  The relatively imprecise
figuring of the large mirror will introduce optical-path errors of
course, but these would be harmless if they are smaller than the
detector time-resolution.  Thus, it appears that a very large
collecting mirror need not result in a loss of resolution.


A further possible application of the same basic idea would be turn
solar power towers into intensity interferometers by night.  Solar
power towers \citep[see e.g.,][]{BREEZE201635} use a large number of
freely-orientable flat mirrors (known as heliostats) to focus light to
a furnace at the top of a tower.  The optical path length is different
for each heliostat, which seemingly precludes interferometry.  But
perhaps the light could be re-collimated by a suitable secondary
mirror such that each heliostat has its own separate sub-beam, which
could go to a detector specific to that heliostat, and then the
detectors could be synchronised in software.  There would still be a
spread in optical paths across each heliostat, thus limiting the
time-resolution $\Delta t$, but perhaps even that could be corrected
for.  But this is very speculative, so we will stop here.

\bigskip\noindent Thanks to Subrata Sarangi for asking all the right
questions that led to this work, and to Nolan Matthews and Nitu Rai
for comments and corrections.

\def\eprint{eprint}
\def\apj{ApJ}
\def\apjl{ApJL}
\def\aap{A\&A}
\def\baas{BAAS}
\def\mnras{MNRAS}
\def\nat{Nature}
\def\pasp{PASP}
\def\procspie{Proc.\ SPIE}

\bibliographystyle{aa}
\bibliography{main.bbl}

\appendix

\section{Rotation of the baseline}\label{sec:matrices}

The $\bb$ vector in this paper is a plane nearly perpendicular to the
line of sight, so that small-angle approximations are valid.  In
general, the $\bb$ plane will not be the horizontal plane
$(x_{\scriptscriptstyle\rm E},x_{\scriptscriptstyle\rm N})$, and a
rotation of coordinates needs to be applied.  The required rotation is
\begin{equation}\label{eq:uvw-rotation}
\begin{pmatrix} x \\ y \\ z \end{pmatrix} =
R_x(\delta) \, R_y(h) \, R_x(-l)
\begin{pmatrix}
x_{\scriptscriptstyle\rm E} \\
x_{\scriptscriptstyle\rm N} \\
x_{\rm up}
\end{pmatrix}
\end{equation}
where
\begin{equation}
R_x(\delta) =
\begin{pmatrix}
1 & 0 & 0 \\
0 & \cos\delta  & -\sin\delta \\
0 & \sin\delta  &  \cos\delta
\end{pmatrix}
\end{equation}
and similarly for $R_x(-l)$ while
\begin{equation}
R_y(h) =
\begin{pmatrix}
 \cos h & 0 & \sin h \\
0       & 1  & 0 \\
-\sin h & 0  & \cos h
\end{pmatrix}
\end{equation}
where $l$ is the latitude of the setup, $\delta$ is the declination
and $h$ is the hour angle of the source.  Expanding out the product
\eqref{eq:uvw-rotation} is equivalent to Eq.~(7) from
\cite{2013APh....43..331D}.  The transformation is basically the same
as in radio-interferometry, with our $(x,y,z)$ being equivalent to
$(u,v,w)$ from radio-astronomy, multiplied by the wavelength.

\section{Numerical implementation}\label{sec:numer}

Given a source brightness $|S(\bO)|$, one can put down a spectrum
$F(\bO,\nu')$ and compute the corresponding $\pha(\bO,t)$ using equation
\eqref{eq:amplV}.  We now see how to implement this numerically.

\subsection{FFT libraries}

Numerical libraries provide efficient implementations of 2D Fourier
transforms.  On an $N\times N$ grid the forward and inverse Fourier
transforms are given as follows.
\begin{equation}\label{eq:fft}
\begin{aligned}
\pha_{pq} =             &\sum_{m,n} \exp\left( 2\pi i\,\frac{pm+qn}N\right)
                                 \, S_{mn} \\
S_{mn} = \frac1{N^2} &\sum_{p,q} \exp\left(-2\pi i\,\frac{pm+qn}N\right)
                                 \, \pha_{pq}
\end{aligned}
\end{equation}
The discrete Parseval relation
\begin{equation}\label{eq:dparseval}
       \sum_{m,n} \left|S_{mn}\right|^2 =
N^{-2} \sum_{p,q} \left|\pha_{pq}\right|^2
\end{equation}
is automatically satisfied.

We could change the definition to put the $1/N^2$ factor in the
forward transform, or have a $1/N$ factor in both the forward and the
inverse transform.  But the above is the standard definition used by
numerical libraries, so let us stay with it.

\subsection{Discretization}

Let us discretize the sky and ground coordinates on grids of size
$N\times N$.  Let
\begin{equation}
\begin{aligned}
\bO &= m\,\Delta s,\, n\,\Delta s \\
\bb &= p\,\Delta b,\, q\,\Delta b \\
\end{aligned}
\end{equation}
where $m,n,p,q$ are all integers in $0,\ldots,N-1$.  The steps $\Delta
s$ and $\Delta b$ are chosen such that
\begin{equation}
\Delta s \times \Delta b = \frac{c/\nu}N
\end{equation}
because then
\begin{equation}
(\nu/c) \, \bb\cdot\bO = \frac{pm+qn}N
\end{equation}
which makes the FFT carry out the Fraunhofer diffraction formula.

The conventional definition \eqref{eq:fft} of the Fourier transform
makes the normalisation of $S_{mn}$ and $\pha_{pq}$ a little tricky.
We can choose $S_{mn}=S(\bO,t)$ or $\pha_{pq}=\pha(\bb,t)$ but not both.
We opt for the latter, and set
\begin{equation}\label{eq:Sgrid}
S_{mn} = \Delta s \;
\frac{\nu/c}{\sqrt{e^{h\nu/(kT(\bO))}-1}}
\; e^{i\rph(\bO)}
\end{equation}
with $T(\bO)$ following the effective temperature map as in
Figure~\ref{fig:src}.  The phase $\rph(\bO)$ we will set below.
Substituting $S_{mn}$ in the discrete Parseval relation
\eqref{eq:dparseval} gives
\begin{equation}
\frac{\nu^2/c^2}{e^{h\nu/(kT(\bO))}-1} \sum_{m,n} (\Delta s)^2
= N^{-2} \sum_{p,q} \left|\pha_{pq}\right|^2
\end{equation}
If we now put
\begin{equation}
\pha_{pq} = \pha(\bb)
\end{equation}
the previous relation has the interpretation
\begin{equation}\label{eq:cparseval}
\int |S(\bO)|^2 d^2\bO = \left\langle |\pha(\bb)|^2 \right\rangle
\end{equation}
which is what we want.  Eq.~\eqref{eq:cparseval} itself is a Parseval
relation corresponding to
\begin{equation}\label{eq:fraunhofer2}
\begin{aligned}
\pha(\bb) = {}&(\nu/c)\, {\textstyle\left(\int d^2\bb\right)^{1/2}} \,\times \\
           &\int e^{2\pi i\,(\nu/c)\,\bb\cdot\bO} \, S(\bO) \, d^2\bO
\end{aligned}
\end{equation}
And now we finally have the full form of Eq.~\eqref{eq:fraunhofer}.
The proportionality factor (which we have not included in
Eq.~\ref{eq:fraunhofer}) is clearly unphysical, but that is just a
consequence of small-angle approximations, and is harmless in
practice.

\subsection{The frequency band}

It remains to set the phases $\rph(\bO)$ in Eq.~\eqref{eq:Sgrid}.  We
let
\begin{equation}
\rph = 2\pi(\nu'-\nu)t
\end{equation}
where $\nu'$ is assigned randomly at each grid-point according to
\begin{equation}
\nu'-\nu = \frac{\tan(\pi r)}{2\pi\Delta\tau} \qquad
-\textstyle{\frac12} < r < \textstyle{\frac12}
\end{equation}
where $r$ is uniformly random.  This distributes $\nu'$ according to
$dr/d\nu'$, which equals the frequency band \eqref{eq:freqfilter} we
wish to implement.  This has $2\pi(\nu'-\nu)$ within
$[-1/\Delta\tau,1/\Delta\tau]$ half the time, but has tails extending
much beyond this range.

\end{document}